\documentclass{emulateapj}
\usepackage{apjfonts}
\usepackage{lscape}
\input{epsf}

\slugcomment{To appear in the Astronomical Journal}

\shorttitle{Four new members of the $\beta$ Pic moving group}
\shortauthors{Lepine \& Simon}

\begin{document}

\title{NEARBY YOUNG STARS SELECTED BY PROPER MOTION. I. FOUR NEW
  MEMBERS OF THE $\beta$ PICTORIS MOVING GROUP FROM THE TYCHO-2
  CATALOG.\altaffilmark{1,2}}

\author{S\'ebastien L\'epine\altaffilmark{3}, \& Michal
  Simon\altaffilmark{4,5}}

\altaffiltext{1}{Based on data obtained in part with the 2.4 m
  Hiltner telescope of the MDM observatory.}

\altaffiltext{2}{Based on data obtained in part with the CTIO
  1.5 m telescope, operated by SMARTS, the Small and Medium
  Aperture Telescope System consortium, under contract with the
  Associated Universities for Research in Astronomy (AURA).}

\altaffiltext{3}{Department of Astrophysics, Division of Physical
Sciences, American Museum of Natural History, Central Park West at
79th Street, New York, NY 10024, USA, lepine@amnh.org}

\altaffiltext{4}{Department of Physics and Astronomy, State University
  of New York, Stony Brook, NY 11794, USA, michal.simon@sunysb.edu}

\altaffiltext{5}{Visiting astronomer, NASA Infrared Telescope Facility
  (IRTF). }

\begin{abstract}
We describe a procedure to identify stars from nearby moving groups and
associations out of catalogs of stars with large proper
motions. We show that from the mean motion vector of a known or
suspected moving group, one can identify additional members of the
group based on proper motion data and photometry in the optical and
infrared, with minimal contamination from background field stars. We
demonstrate this technique by conducting a search for low-mass members
of the $\beta$ Pictoris moving group in the Tycho-2 catalog. All known
members of the moving group are easily recovered, and a list of 51
possible candidates is generated. Moving group membership is evaluated
for 33 candidates based on X-ray flux from $ROSAT$, H$\alpha$ line
emission, and radial velocity measurement from high-resolution
infrared spectra obtained at $Infrared Telescope Facility$. We confirm
three of the candidates to be new members of the group: TYC
1186-706-1, TYC 7443-1102-1, and TYC 2211-1309-1 which are late-K and
early-M dwarfs 45\--60pc from the Sun. We also identify a common
proper motion companion to the known $\beta$ Pictoris Moving Group
member TYC 7443-1102-1, at a 26$\arcsec$.3 separation; the new
companion is associated with the X-ray source 1RXS
J195602.8-320720. We argue that the present technique could be applied
to other large proper motion catalogs to identify most of the elusive,
low-mass members of known nearby moving groups and associations.

\end{abstract}

\keywords{astrometry \--- stars: emission-line, Be \--- stars:
  kinematics \--- stars: pre-main sequence \--- open clusters and
  associations: individual ($\beta$ Pictoris Moving Group) }

\section{Introduction}

Young stars are critical in understanding not just the star-formation
process itself but also the formation of planetary systems, through
observation of their circumstellar environment
\citep{JHFFTP99,MHM04,LMWK04,ZS04b,WBZS04,Cetal05}. This is most
efficiently achieved from the nearest possible vantage point, which
makes nearby young stars very sought-after targets
\citep{G98,Montes01,WCH03,King03,ZS04,TQMS08}. Nearby young stars have
become prime targets to achieve direct exoplanet imaging
\citep{Neu03,MZ04,MMHAB05,Low05,Laf07,DSRC07,KAJB07} because massive planets
are expected to be much more easily detected in their youth at a time
when they still shine bright from their own gravitational collapse.

Young stars are known to exist in the vicinity of the Sun as members
of ``moving groups'', each consisting in loose associations of stars
moving in the same approximate direction within the solar
neighborhood. Proposed nearby moving groups and associations include
the Castor moving group \citep{B98,R03}, the $\beta$ Pictoris moving
group \citep{Z01}, the Tucana/Horologium association \citep{SZB03},
the AB Doradus moving group \citep{ZSB04}, the Carina-Near moving
group \citep{ZBSK06}, and notably the very young TW Hydrae association
\citep{WZPPWSM99,SACP99,SZB03} whose probable members include several
brown dwarfs \citep{G02,SMZL05,LBKS07}. All of these nearby moving
groups and associations are suspected to be dispersed fragments from a
larger neighboring OB association \citep{BB02,O04,M07}.

Stars from nearby moving groups and associations are identified and
confirmed through a combination of kinematic data and evidence of
youth. Most of the original members were identified based on their
full (three dimentional) kinematics provided by accurate proper motion
and parallax from the Hipparcos catalog \citep{P97}, complemented with
radial
velocity observations. While it is relatively straightforward to
identify moving group members among samples of stars for which both
proper motion {\em and} parallax data exist, it is more complicated to
search for members among field stars with no parallax data in
hand. Proper motion measurements are now widely available for vast
numbers of stars from all-sky proper motion catalogs such as Tycho-2
\citep{H00}, USNO-B1.0 \citep{M03}, or SUPERBLINK \citep{LS05}, but
parallax and radial velocity measurements remain relatively rare,
particularly for stars with visual magnitude $V\gtrsim10$ mag. As a
result, one must rely on secondary diagnostics, such as unambiguous
signs of youth, to search for moving group members. Young low-mass
stars are expected to have faster rates of rotation and thus higher
coronal activity through a stronger dynamo effect
\citep{MLFG01,K03}. High coronal activity is associated with strong
emission lines of atomic hydrogen and large X-ray
emission. Spectroscopic follow-up of X-ray sources from the {\em
  ROSAT} catalogs has yielded positive identification of many nearby
young stars, including new members of the TW Hydrae association
\citep{TNFG03} and the $\beta$ Pictoris moving group
\citep{TQSRMS06}. However, the {\it ROSAT} catalogs are flux limited, and
though they are useful in identifying nearby young stars, they cannot
serve as the basis for a {\em complete census} of moving group
members.

It is telling that most confirmed members of young moving groups
consist in stars of spectral subtype F, G, or K, and include few
low-mass stars of spectral type M. Assuming those groups to follow the
standard initial mass function, one would expect them to have
significantly larger numbers of associated low-mass stars. Such
low-mass members arguably are particularly promising targets for
exoplanet surveys \citep{LS06} because the detectability of planets of
a given mass is improved when the parent star has low intrinsic
luminosity.

In this paper, we develop a general technique to identify new members
of known moving groups among stars listed in large proper motion
catalogs. We demonstrate the technique by performing a search for new
members of the $\beta$ Pictoris moving group (BPMG), using a subsample
of stars from the Tycho-2 catalog \citep{H00} with proper motions
$\mu>70$ mas yr$^{-1}$. The general method is described in Section 2. A list
of candidates members of the $\beta$ Pictoris moving group is
identified in Section 3. Follow-up spectroscopic observations are described
in Section 4, and new $\beta$ Pictoris members are identified. Conclusions
follow in Section 5.

\begin{figure*}
\epsscale{1.15}
\plotone{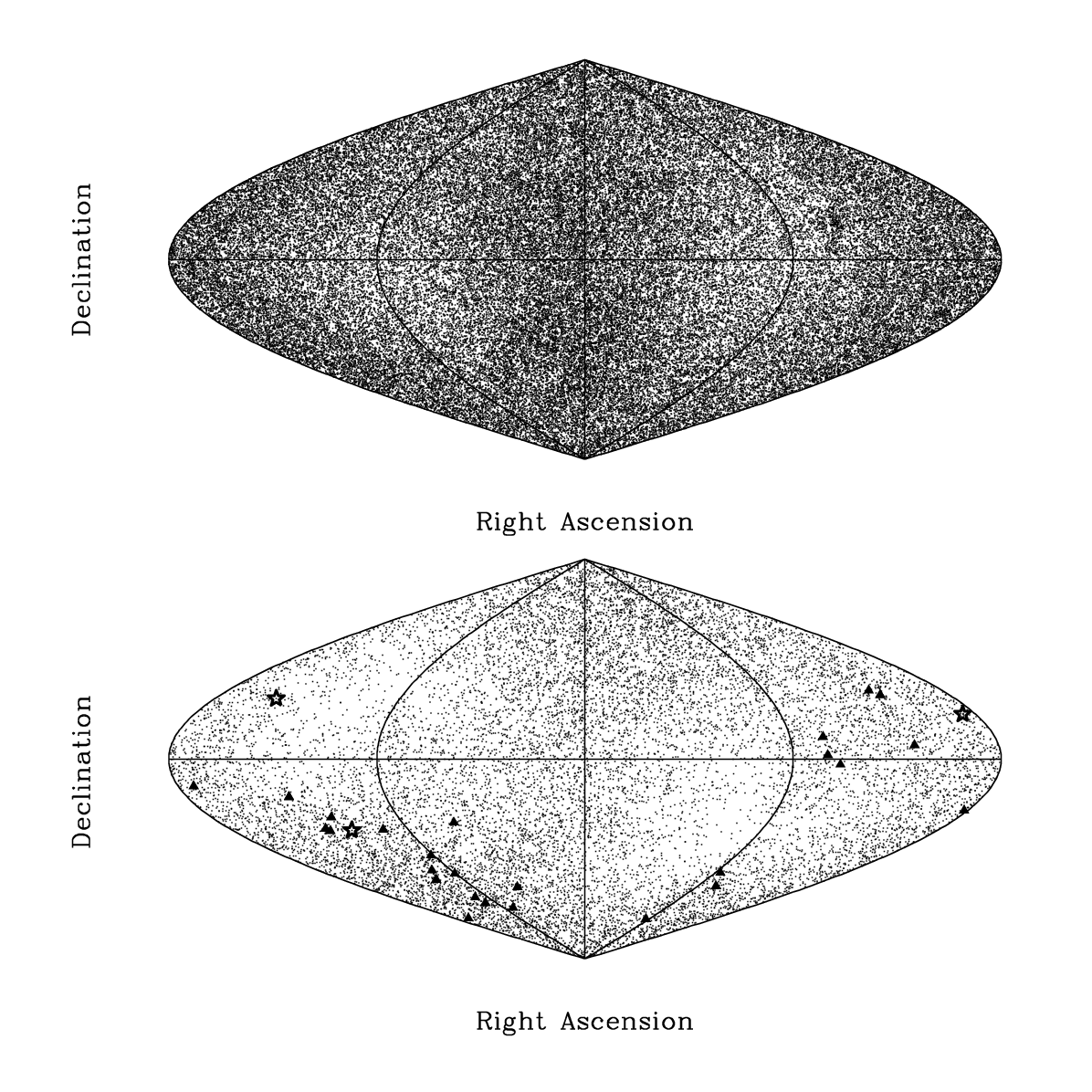}
\caption{Top: distribution on the sky of the 86,626 stars in the
  Tycho-2 catalog which have proper motions $\mu>70$ mas yr$^{-1}$,
  each dot
  representing a star. Note the nonuniform distribution which is a
  result of the Sun's systemic motion through the local standard
  of rest and the systemic drift of thick disk and halo stars. Bottom:
  the subsample of 15,989 stars whose proper motion angle is
  consistent with membership in the $\beta$ Pictoris moving
  group (proper motion angle within $12^{\circ}$ of the local
  projected vector of the mean $\beta$ Pictoris group
  motion). 29 known $\beta$ Pictoris members are recovered
  in the process (triangles). Open star symbols show the locations
  of the three new confirmed member systems.}
\end{figure*}

\section{Proper motion selection method}

Let $\vec{V}_{\rm mg}= U_{\rm mg} \hat{x} + V_{\rm mg} \hat{y} +
W_{\rm mg} \hat{z}$ be the mean motion of a local moving group, in the
local standard of rest, with $\hat{x}$ a unit vector pointing in the
direction of the Galactic center, $\hat{y}$ in the direction of the
Sun's Galactic orbital motion, and $\hat{z}$ pointing toward the
north Galactic pole. At any point on the sky described by the Galactic
coordinates $(l,b)$, one can calculate the local projected motion in
the plane of the sky $({v_l}_{\rm mg},{v_b}_{\rm mg})$ of the moving
group in the direction of Galactic longitude $l$ and Galactic latitude
$b$ from
\begin{equation}
{v_l}_{\rm mg} = \-- U_{\rm mg} \sin{(l)} + V_{\rm mg} \cos{(l)}
\end{equation}
\begin{equation}
{v_b}_{\rm mg} = \-- U_{\rm mg} \cos{(l)} \sin{(b)} \-- V_{\rm mg}
\sin{(l)} \sin{(b)} + W_{\rm mg} \cos{(b)}.
\end{equation}

In Galactic coordinates, the angle $\Phi$ that this vector
$({v_l}_{\rm mg},{v_b}_{\rm mg})$ subtends with a star's local
proper motion vector $(\mu_l,\mu_b)$ is
\begin{equation}
\cos(\Phi) = \frac{\mu_l {v_l}_{\rm mg} +  \mu_b {v_b}_{\rm
  mg}}{({\mu_l}^2 +  {\mu_b}^2)^{\frac{1}{2}} \ ( {{v_l}_{\rm mg}}^2 +
  {{v_b}_{\rm mg}}^2)^{\frac{1}{2}}} ,
\end{equation}
wich is easily derived from the definition of the scalar product
of the two vectors. Stars which are actual members of the moving
groups will have their proper motion vector closely aligned with the
projected motion of the group, and will thus have
$\cos(\Phi)\simeq1.0$. One can thus sift through a proper motion
catalog and search for potential members of the moving groups by
considering only the stars whose proper motion vector is within a few
degrees of the expected orientation of the projected motion of the
moving group. This restriction can be expressed simply as:
\begin{equation}
cos(\Phi) > \cos(\Phi_{\rm max}) \equiv Z_{\rm max} .
\end{equation}
The particular choice of $\Phi_{\rm max}$ (or $Z_{\rm max}$) will
depend on the accuracy of the catalog's proper motions, and also on
the dispersion in velocities between stars of the moving group. For
example, $Z_{\rm max}=0.985$ would restrict the sample only to stars
whose proper motion is within $\approx10^{\circ}$ of the expected
projected motion of the moving group. A value of $Z_{max}$ closer to
$1.0$ will considerably restrict the search, but may overlook some
actual members. A smaller value of $Z_{max}$ will be more likely to
include all members of the group, but will also produce a larger
sample with more contaminants.

Proper motion aligned in the direction of the projected motion of
the moving group is however not a sufficient condition for a star to
be a member of the group, because a given proper motion vector may
correspond to a range of transverse velocities, depending on the
star's distance. If the distance $d$ to the star is
known, then an actual member of the moving group will also verify:
\begin{equation}
\mu = ({\mu_l}^2 + {\mu_b}^2)^{\frac{1}{2}} \simeq 0.211 \ d^{-1}
( {{v_l}_{\rm mg}}^2 + {{v_b}_{\rm mg}}^2)^{\frac{1}{2}} ,
\end{equation}
with ${v_l}_{\rm mg}$ and ${v_b}_{\rm mg}$ given by Eqs.1-2
above, and where the velocities are in km s$^{-1}$ and the distance $d$
in parsecs. However, distances are generally not known for stars in
proper motion catalogs, the one notable exception being the {\it Hipparcos}
catalog. For a star with no recorded distance, it is however possible
to constrain the distance based on color\--magnitude relationships. If
one assumes that a star is a member of the moving group, then it is
possible to get an accurate distance estimate for that star based on
the magnitude of the proper motion alone; this hypothetical distance
can then be checked for consistency using color\--magnitude
relationships. Assuming that one given proper motion star is also a member
of the moving group, then its hypothetical, {\em kinematically
  derived} distance $d_{\rm kin}$ will be
\begin{equation}
d_{\rm kin} \simeq 0.211 ( {{v_l}_{\rm mg}}^2 + {{v_b}_{\rm
    mg}}^2)^{\frac{1}{2}} ( {{\mu_l}}^2 + {\mu_b}^2)^{-\frac{1}{2}} .
\end{equation}
If $d_{\rm kin}$ is found to be inconsistent with some other distance
estimate, such as a photometric distance, a spectroscopic distance or
better yet a parallax distance, then the star can be ruled out as a
possible member of the moving group.

Unless parallax measurements are already available for the candidate
star, it is useful to compare $d_{\rm kin}$ with the photometric
distance. Consider the visual magnitude $V$ of a proper motion star
and its infrared $K$ magnitude, if the object is a main-sequence star,
then it can be expected to follow a color\--magnitude relationship
$M_K=M_K(V-K)$ whose form depends on the age of the star. If
a star is a member of the moving group under consideration, and
assuming interstellar extinction to be negligible (generally valid for
nearby stars), then one expects
\begin{equation}
K + 5 \log{(d_{\rm kin})} + 5 - M_K(V-K) \lesssim \sigma_K ,
\end{equation}
where $\sigma_K$ is the mean dispersion about the color\--magnitude
relationship for members of that particular moving group.

An additional test is to verify that the radial velocity of the
prospective moving group member is also consistent with group
membership. A moving group member can be predicted to have a radial
velocity ${v_r}_{\rm mg}$ such that 
\begin{equation}
{v_r}_{\rm mg} = + U_{\rm mg} \cos{(l)} \cos{(b)} + V_{\rm mg}
\sin{(l)} \cos{(b)} + W_{\rm mg} \sin{(b)} .
\end{equation}
This prediction can be tested with radial velocity
measurements. 

Ultimately, one would want to confirm group membership by accurately
measuring the $UVW$ components of velocity of the source, which
requires accurate distance measurement through geometric parallax as
well as accurate radial velocities from high-resolution
spectroscopy \citep{MLFG01}. Additionally, moderate- to
high-resolution spectroscopy should be used to check for the presence
of lithium in the atmosphere (\ion{Li}{1} $\lambda$6708 in emission) which is
a strong diagnostic for young age as Li is progressively depleted over
time in low-mass stars \citep{Michaud91}.

However such a detailed analysis of candidate members is
expensive. High-resolution spectroscopy of moderately faint stars
requires long observing times and/or large telescope
apertures, and parallax measurements are intensive and require years
of careful monitoring. It would be desirable to pre-select stars with
the highest predicted likelihood of being moving group members. Such
stars are expected to display secondary evidence of youth, which
includes evidence for atmospheric activity such as X-ray flux or
H$\alpha$ line emission. While activity in itself does not necessarily
imply that a star is young, a high proportion of the most active stars
are found to be young objects \citep{Jeffries1995}.

The above equations and practical considerations provide a road map
for identifying elusive members of a known moving group out of a
statistically complete catalog of stars with large proper motion. 
\begin{enumerate}
\item{Isolate a subsample of high proper motion stars whose proper
  motion vector points in the expected direction for the moving group
  (Eq. 4).}
\item{Identify stars from that subsample for which photometric data
  shows a match between the hypothetical kinematic distance and the
  photometric distance (Eq.7).}
\item{Reduce the sample further by selecting targets showing evidence
  for activity, such as strong X-ray or H$\alpha$ emission.}
\item{Compare the predicted radial velocity (Eq.8) to radial velocity
  measurements to identify probable kinematic members of the group.}
\item{Confirm kinematic membership through parallax measurements}
\item{Confirm young age through detection of \ion{Li}{1} $\lambda$6707.}
\end{enumerate}
In Section 3 below, we provide a demonstration of steps 1-4, applied to a
subsample of bright high proper motion stars.

\section{A search for $\beta$ Pictoris moving group members in the
  Tycho-2 catalog}

We demonstrate the technique described above by performing a search
for members of the $\beta$ Pictoris moving group in a subsample of high
proper motion stars from the Tycho-2 catalog of \citet{H00}. We have
assembled a list of 86,626 stars with proper motions $\mu>70$ mas
yr$^{-1}$; their distribution on the sky is shown in Figure 1 ({\it
  top panel}). This represents an all-sky sample of relatively bright
stars. All objects are listed with Tycho-2 visual magnitudes
$V_T<14.0$ mag, with $90.7\%$ of the stars in the $V_T<12.0$ mag
range. We have used the VizieR
service\footnote{http://webviz.u-strasbg.fr/viz-bin/VizieR} and
additional software of our own to cross-correlate the entire list with
the Two Micron All-Sky Survey (2MASS) All-Sky Catalog of Point Sources
\citep{C03} and obtain infrared $J$, $H$, and $K_{\rm s}$ magnitudes
for all the stars (a 10$\arcsec$ search radius was used, and proper
motions taken into account; most counterparts were recovered within
2$\arcsec$ of their extrapolated position). 


\subsection{Proper motion selection}

The first selection cut selects for stars which have a proper motion
vector whose orientation is consistent with BPMG membership. We adopt
as the fiducial space motion of the BPMG the mean motion of its known
members, estimated by \citet{TQSRMS06} to be $(U_{\rm mg}, V_{\rm mg},
W_{\rm mg})=(-10.1, -15.9, -9.1)$ km s$^{-1}$ relative to the Sun. We apply
Eq.~3, and select Tycho-2 stars whose proper motion vector makes an
angle $\Phi<12^{\circ}$ with the local projected BPMG motion
vector ($Z_{max}$=0.978). This reduces the sample to 15,989 stars
(18.4\% of the initial sample) whose proper motion vector is aligned
with the local projected motion of the BPMG. Their distribution on the
sky is shown in Figure 1 ({\it  bottom panel}). The distribution 
shows a dipolar moment, whose axis is aligned along the apex of the
BPMG motion. 

We have compared our list of proper motion selected objects to the 51
known members of the BPMG listed in \citet{TQSRMS06} and
\citet{TQMS08}. We find that we recover 30 of the known members. Of the
22 stars that are not recovered, we find nine stars that are too
faint to be listed in the Tycho-2 catalog (8 of them being faint
companions of recovered group members). Ten other stars have proper
motions smaller than the proper motion limit of our initial sample
($\mu>70$ mas yr$^{-1}$) and thus could also not possibly have been
recovered. The two remaining stars are bright high proper motion
stars which are listed in the {\it Hipparcos} catalog but, for some reason,
are missing from the Tycho-2.\footnote{The two Hipparcos stars missing
  from the Tycho-2 catalog are HIP 12545 and HIP 112312.} All stars
being accounted for, we then find our proper motion selection
algorithm to have a recovery rate of 100\%. This suggests that any
unrecognized groups member that is present in the initial sample will
be selected in the proper motion cut.

The 30 recovered members are listed in Table 1, and their location on
the sky is indicated in Figure 1 with solid triangles. A majority of the
members are located in the southern sky, with a concentration around
$(\alpha,\delta)\sim(270^{\circ},-45^{\circ}$). However this moving
group extends over a broad swath of the sky, and even has a few
members north of the celestial equator. It even is possible that the
Sun could be situated within the confines of the group. Hence, any of
the $15,962$ other stars could potentially be a $\beta$ Pictoris
member, regardless of its position on the sky.

\subsection{Kinematic distance selection}

The largest reduction in the number of possible BPMG candidates is
achieved through kinematic distance selection. Based on data from the
{\it Hipparcos} catalog, we calculate that the median transverse velocity of
stars with proper motion $\mu>70$ mas yr$^{-1}$ is $\simeq47$ km
s$^{-1}$. The mean motion of the BPMG is only $29$ km s$^{-1}$, which
means that field stars are moving faster on average, and thus $d_{\rm
  kin}$ will tend to underestimate their distances, making field stars
appear underluminous.

\begin{figure}
\epsscale{1.2}
\plotone{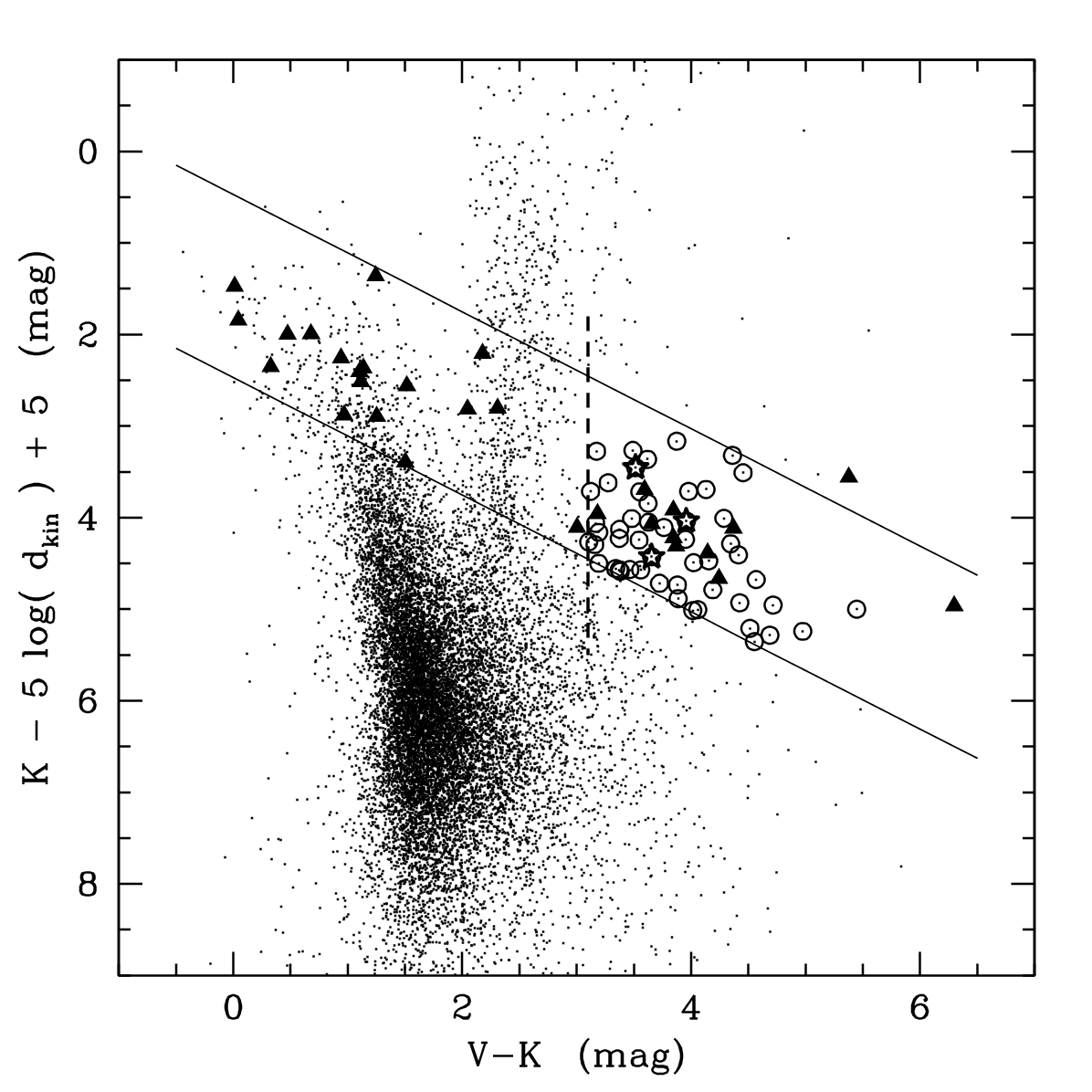}
\caption{Color\--magnitude diagram of the 15,989 stars with proper
  motions angles consistent with $\beta$ Pictoris group membership, with
  kinematic distances ($d_{\rm kin}$) calculated under the assumption that
  all stars are group members. The 30 known group members (triangles)
  are aligned along the expected main sequence for young stars of that
  age. Stars which do not fall close to that locus have inconsistent
  $d_{\rm kin}$ values, and can be rejected as unlikely group members. A
  high level of contamination from distant giants is observed around
  $V_T-K_s\simeq2.5$. We have identified 51 very red objects
  ($V_T-K_s>3.1$) as low-mass stars with a high potential for group
  membership (open symbols). 33 are investigated further in
  this paper; the three candidates confirmed in this paper to be very
  likely group members are plotted with five-pointed star symbols.}
\end{figure}

Figure 2 shows the color\--magnitude diagram for all 15,989 Tycho-2
stars identified as possible members, with $d_{\rm kin}$ used to
calculate a pseudo absolute magnitude. The figure reveals that known
group members occupy a distinct locus in this color\--magnitude diagram,
with most members following a relatively well defined color\--magnitude
relationship. This is not unexpected, as the calculated $d_{\rm kin}$
should be very close to the actual distance $d$ to those stars. We
find that one can model the mean locus of the BPMG members with
\begin{equation}
K_s - 5 log(d_{\rm kin}) + 5 = 1.6 + 0.65 ( V_T - K_s ) 
\end{equation}
with all members but one falling within $\pm1$ mag of this line.

As expected, the vast majority of stars in the overall sample do
not cluster along the line defined in Eq.9, but appear
significantly underluminous in the color\--magnitude
diagram. Nonmembers may fall off the well-defined locus of the group
members for two main reasons: (1) their calculated $d_{\rm kin}$
either underestimate or overestimate their true distance, and/or (2)
the stars are significantly older than BPMG members, and do not follow
the same color\--magnitude relationship as the young BPMG stars. From
Figure 2, we see that most stars fall well below the standard
color\--magnitude relationship of BPMG members, most probably because
$d_{\rm kin}$ significantly underestimate their true distance. All
these background stars can be efficiently eliminated from the list of
prospective candidates.

We find that of the initial 15,962 prospective members, only 835 fall
within $\pm1$ mag of the mean color\--magnitude relationship of BPMG
members. This reduces the list of possible candidates by a factor of
20. The much reduced subsample, however, remains contaminated by a
significant number of background giants which are easily identified
in the color\--magnitude diagram. These appear to be red clump K
giants, with optical-infrared colors 2$\lesssim V_T-K_s\lesssim$3. In
Figure 2, they show up as a nearly vertical band of stars extending all
the way up the diagram. These stars are a serious source of
contaminants, however they occupy a fairly narrow range of
colors. Since we are mostly interested in identifying the elusive
low-mass members of the BPMG, we can further restrict our search to
stars with very red colors, thus eliminating most of those K
giants. We introduce $V_T-K_s>3.1$ as additional color cut (dashed
line in Figure 2. This further reduces the sample to a short list of
55 probable {\em low-mass} BPMG members. Such a sample is of a size
very much manageable for follow-up spectroscopic programs. 

A search of the literature reveals that four of the 55 candidates
are actually known members of the TW Hya association (TWA). The
reason for their presence among BPMG candidates is due to the fact
that the TWA mean motion vector
$(U,V,W)_{TWA}\simeq(-11.0,-18.0,-5.0)$ km s$^{-1}$ \citep{ZS04} is
very close to the mean motion vector of the BPMG
$(U,V,W)_{BPMG}\simeq(-10.1,-15.9,-9.1)$ km s$^{-1}$
\citep{TQSRMS06}. In many directions on the sky, the two associations
run nearly parallel to each other, which makes confusion
possible. Both groups also contain young stars, and occupy similar
loci in the color\--magnitude diagram. The four TWA interlopers are
listed in Table 1. 

Of the 51 remaining candidates, we proceeded to investigate further 33
objects which were within range for our follow-up spectroscopic
observations (see below). This short list of 33 candidates is provided
in Table 2. Follow-up observations are now being planned for the
remaining 18 targets, which will be presented in a separate paper.

\section{Confirmation of new BPMG members}

\subsection{Hipparcos parallaxes}

Table 2 compiles data on the 33 candidates followed up in this
paper. The table lists the Tycho-2 catalog (TYC) identifier, HIP
numbers for stars listed in the {\it Hipparcos} catalog, coordinates in the
ICRS system, proper motions, Tycho-2 visual magnitude V$_{\rm T}$ and
2MASS infrared K$_{\rm s}$ magnitude. The table also lists the
predicted distances assuming BPMG membership ($d_{\rm kin}$, based on
the proper motion). 

\begin{figure}
\epsscale{1.2}
\plotone{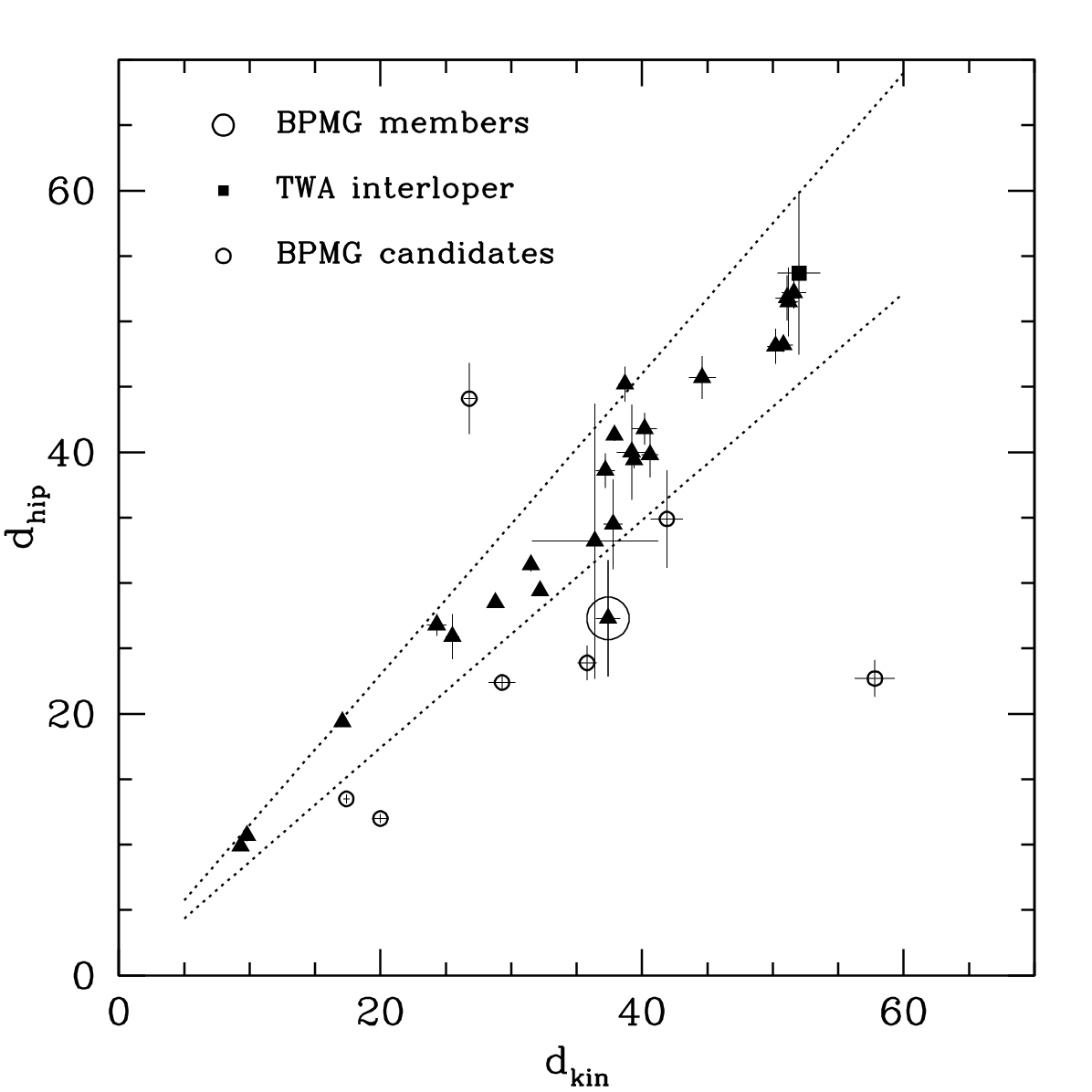}
\caption{Comparison between the kinematic distance $d_{\rm kin}$ of
  recovered and prospective members of the BPMG, and their distances
  based on Hipparcos parallax $d_{\rm hip}$. All previously known members
  (triangles) have both quantities agreeing to within 15\% except for
  BD+28$^{\circ}$382B (circled) whose distance is underestimated by
  Hipparcos (see the text). The solid square shows the location of the one
  TWA interloper with Hipparcos parallax (it is TW Hya itself).}
\end{figure}

Though parallaxes are not generally available for stars in the Tycho-2
catalog, most of the brighter (V$_{\rm T}<10$) entries are also
{\it Hipparcos} stars, and accurate parallaxes exist for them. In Tables 1
and 2, we identify all the {\it Hipparcos} objects by their catalog number,
and provide distances (d$_{\rm hip}$) based on their {\it Hipparcos}
parallaxes as obtained from the recent new reduction of
\citet{vL07}. Distances derived from these parallaxes are also
tabulated; they are generally accurate to $\pm$1-2 parsecs, though a
few have large uncertainties, including TYC 689-130-3 (=[RHG95]853, =LP
476-207) whose estimated distance is 33.2$\pm$10.5 pc. Additionally, the
parallax to BD+28$^{\circ}$382B, the resolved companion to HD 14082,
is clearly overestimated in the \citet{vL07} catalog, as it is quoted
to be 36.58$\pm$5.83 mas, compared with 28.97$\pm$2.88 mas for HD
14082.

Figure 3 plots d$_{\rm hip}$ against the kinematic distance d$_{\rm
 kin}$, calculated from the proper motion assuming all stars to be
moving with the mean motion of the $\beta$-Pictoris moving group
 (Equation 6). We provide uncertainties on d$_{\rm kin}$ which are based on
the uncertainties in the Tycho-2 proper motions. The relative errors
on the proper motions are systematically smaller than those on the
parallaxes, which yield smaller errors on d$_{\rm kin}$ than on
d$_{\rm hip}$. One should keep in mind, however, that the d$_{\rm
kin}$ assume the stars to be moving exactly at the mean motion of the
group, which may not be strictly accurate because group members have
some dispersion about the mean velocity. We find that all confirmed
BPMG members have d$_{\rm hip}$ within 20\% of d$_{\rm kin}$. The
small scatter is due to parallax and proper motion accuracies, but
also to the intrinsic velocity dispersion of individual group members
which is unaccounted in d$_{\rm kin}$. 

Of the 33 BPMG candidates, we find seven {\it Hipparcos} stars whose parallax
distances (d$_{\rm hip}$) are significantly different from the
predicted kinematic distances (d$_{\rm kin}$); this eliminates them as
possible group members. Six have d$_{\rm kin}$ overestimating their
true distance; these must be background stars with transverse velocities
larger than the BPMG. The other star has d$_{\rm kin}$ underestimating its
true distance, in this case the star must be a foreground object moving
slower than the BPMG. With those eliminated from the sample, our
follow-up observations have focused on the remaining 26 candidates.

\subsection{Stellar activity: X-ray emission}

We searched for counterparts to the BPMG candidates in the {\it ROSAT}
X-ray catalogs, both the {\it ROSAT} All-Sky Bright Source Catalog of
\citet{Voges99}, and the {\it ROSAT} All-Sky Survey Faint Source Catalog
\citep{V00}. We also searched for counterparts to the known BPMG
members. According to \citet{Voges99}, the positional accuracy of the
{\it ROSAT} catalog for point sources is 13$\arcsec$, 90\% of the time, with
some outliers possibly having errors as large as 40$\arcsec$. To make
sure we did not overlook possible X-ray counterparts, we have
examined all {\it ROSAT} sources within 50$\arcsec$ of our BPMG candidates.

We found possible X-ray counterparts for 22 of the known BPMG
members, and also to all four of the TW Hya
association interlopers. We further found counterparts to six of the
BPMG candidates. Of the 32 possible X-ray counterparts, 25 have {\it ROSAT}
positions placing them within 15$\arcsec$ of the Tycho-2 star, and
are thus convincing matches. The other seven sources have large
offsets, and were investigated further. A counterpart to
TYC 6412-1068-1 (=HD 203), the source 1RX J000648.9-230608, is found
with an offset of 24$\arcsec$. An examination of scans from the Digitized Sky
Surveys (DSS) and from 2MASS did not reveal any optical of infrared
source within 15$\arcsec$ of the quoted {\it ROSAT} position. Furthermore,
the {\it ROSAT} catalog reports a positional error of 19$\arcsec$ for that
X-ray source. Given the larger positional error and the absence of any
alternative, we conclude that 1RX J000648.9-230608 is an actual
counterpart of TYC 6412-1068-1 (=HD 203). Likewise the {\it ROSAT} sources 1RXS
J184523.4-645201 and 1RXS J184657.3-621037 were found to be near the
stars TYC 9077-2487-1 (=HD 172555) and TYC 9073-762-1, with offsets of
only 26$\arcsec$ and 33$\arcsec$, respectively. The two sources are
quoted to have {\it ROSAT} positional errors of 21$\arcsec$ and 24$\arcsec$,
and we could not find any other possible match to an optical or
infrared source. We thus identify these sources as the actual X-ray
counterparts to TYC 9077-2487-1 and TYC 9073-762-1. The star
TYC 7760-283-1 has a possible X-ray counterpart (1RXS J121527.9-394843)
whose {\it ROSAT} position is 33$\arcsec$ from the Tycho-2 position. The quoted
{\it ROSAT} positional uncertainty is 14$\arcsec$, which would suggest the
X-ray source is not a match. However, examination of the DSS and 2MASS
reveals no other possible optical counterpart. Because the hardness
ratios are consistent with the other BPMG and TW Hya association
members, we assume the X-ray source to be a match, although more
accurate X-ray astrometry would have to confirm this. Among the new
BPMG candidates, we found the star TYC 643-73-1 to have an associated
X-ray source (1RXS J024419.4+105707) with a {\it ROSAT} position off by
45$\arcsec$. Again the positional uncertainty quoted in the {\it ROSAT}
catalog is relatively large (43$\arcsec$) and we thus conclude that
the two are the same object; no other possible counterpart can be seen
in the DSS and 2MASS images. The X-ray count rates of all identified
counterparts are listed in Tables 1 and 2. We also give the
two {\it ROSAT} hardness ratios $HR1$ and $HR2$.

Finally, the star TYC 7443-1102-1 is found to be 26$\arcsec$ from a
{\it ROSAT} source (1RXS J195602.8-320720) which has a quoted positional accuracy
of only 9$\arcsec$. The source is however not a counterpart to
TYC 7443-1102-1.\footnote{We thank the referee, J. A. Caballero, for pointing
 this out.} A close examination of the DSS images reveals that the
X-ray source is a near-perfect match to a nearby, fainter
star. Interestingly, a search of the USNO-B1.0 reveals that this star
has a proper motion [$\mu RA,\mu Decl$]=[30,-78] mas yr$^{-1}$ which is
nearly coincident with the proper motion of TYC 7443-1102-1,
[$\mu RA,\mu Decl$]=[30,-66]. The two stars appear to form a common
proper motion pair. Based on the USNO-B astrometry, the companion is
at an angular separation $\rho$=26$\arcsec$.3 from TYC 7443-1102-1,
with a position angle $\theta$=316.5 degrees. Assuming TYC 7443-1102-1 to
be a member of the BPMG, the kinematic distance of 45.6$\pm$1.6
parsecs suggests a wide binary with projected orbital separation of
1,200$\pm$42 AU.

To summarize, we find that the majority (21 of 30) of the known BPMG
members have X-ray counterparts in the {\it ROSAT} catalogs, which indicates
chromospheric activity consistent with their young ages. Likewise, all
four TWA interlopers have bright associated {\it ROSAT} counterparts. On the
other hand, we find X-ray counterparts for only seven of the 33 candidates
under investigation (Table 2). This suggests that a significant fraction
of the candidates are not young/active and are thus not genuine BPMG
members but field stars with projected motions aligned with the
projected BPMG velocity vector.

Interestingly, the counterparts of the BPMG candidates have hardness
ratios consistent with those of moderately young stars in nearby
moving groups. The counterparts are distributed on a locus which
coincides roughly with the locus of the known BPMG members (Figure
4). All of them in turn have hardness ratios intermediate between
those of extremely young T Tauri stars, and those of the older field K
and M dwarfs \citep{K03}. This suggests that BPMG candidates with
X-ray counterparts are likely to be nearby stars of relatively young
ages, whether or not they are actual BPMG members.

Because some of the known BPMG members also do not have counterparts
in {\it ROSAT}, we find that the absence of a {\it ROSAT} counterpart is not
sufficient to justify rejection. We thus complement the {\it ROSAT} data
with a spectroscopic search for atomic line emission.

\begin{figure}
\epsscale{1.2}
\plotone{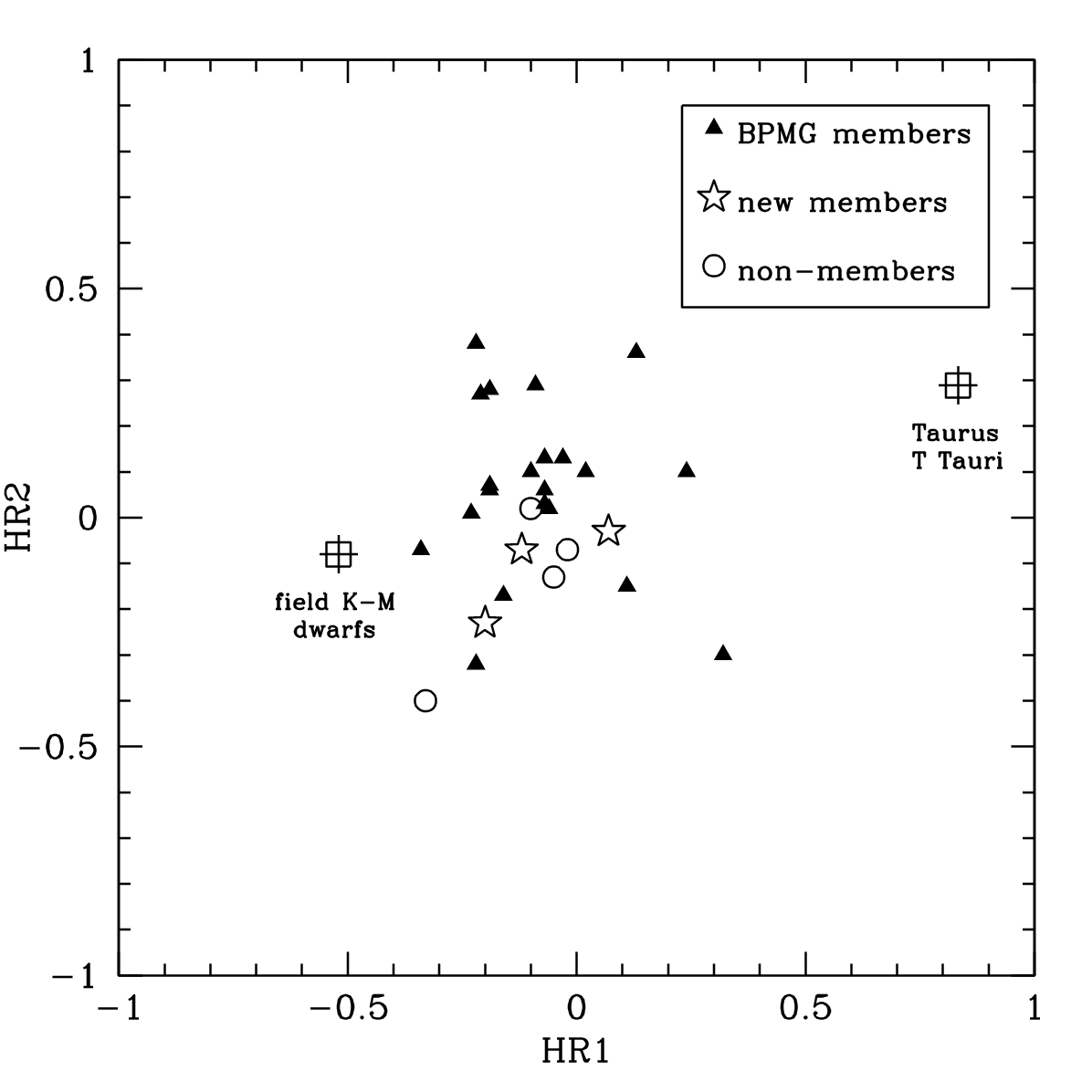}
\caption{X-ray hardness ratio $HR1$ vs. hardness ratio $HR2$ for the
  stars investigated in this paper. Both $HR1$ and $HR2$ are from the {\it ROSAT}
  All-Sky Catalog. The candidates (open symbols) occupy a locus
  coincident with the known BPMG members (filled triangles),
  intermediate between the mean locus of older field K and M dwarfs,
  and the locus of very young T Tauri stars \--- both values from
  \citet{K03} \--- which have the strongest hardness ratios. The
  uncertainties for the values from  \citet{K03} are standard errors
  of the mean; the actual scatter of the individual points is
  comparable to that of the BPMG members plotted here.}
\end{figure}

\subsection{Stellar activity: H$\alpha$ emission}

We have carried out a follow-up program of low-resolution
spectroscopy to measure the strength of the H$\alpha$ emission line in
BPMG candidates. Targets in the northern sky were observed from MDM
observatory on Kitt Peak, with the 2.4 m Hiltner
telescope. Spectra were obtained with the MkIII spectrograph, using
the thick frontside-illuminated 2K$\times$2K Loral CCD camera
(``Wilbur''), which has a pixel size of 15$\mu$m. We used the 300
lines mm$^{-1}$ grating blazed at 8000\AA\ in first order to produce spectra
with a resolution of 6.0\AA. With the telescope at f/7.5
focal, the spatial scale was 0$\arcsec$.878 pixel$^{-1}$. The
1\arcsec.5 wide, longslit spectra were reduced using IRAF, following
standard flatfield correction, sky background subtraction, and
spectrum extraction. The wavelength scale was calibrated using
an arc spectrum of NeAr. Flux calibration was determined from
same-night observations of a set of spectrophotometric standards
(Feige 66, Feige 110, Wolf 1346) from \citet{Oke90}. Effective
spectral coverage is 5800\AA-8300\AA.

\begin{figure}
\epsscale{2.6}
\plotone{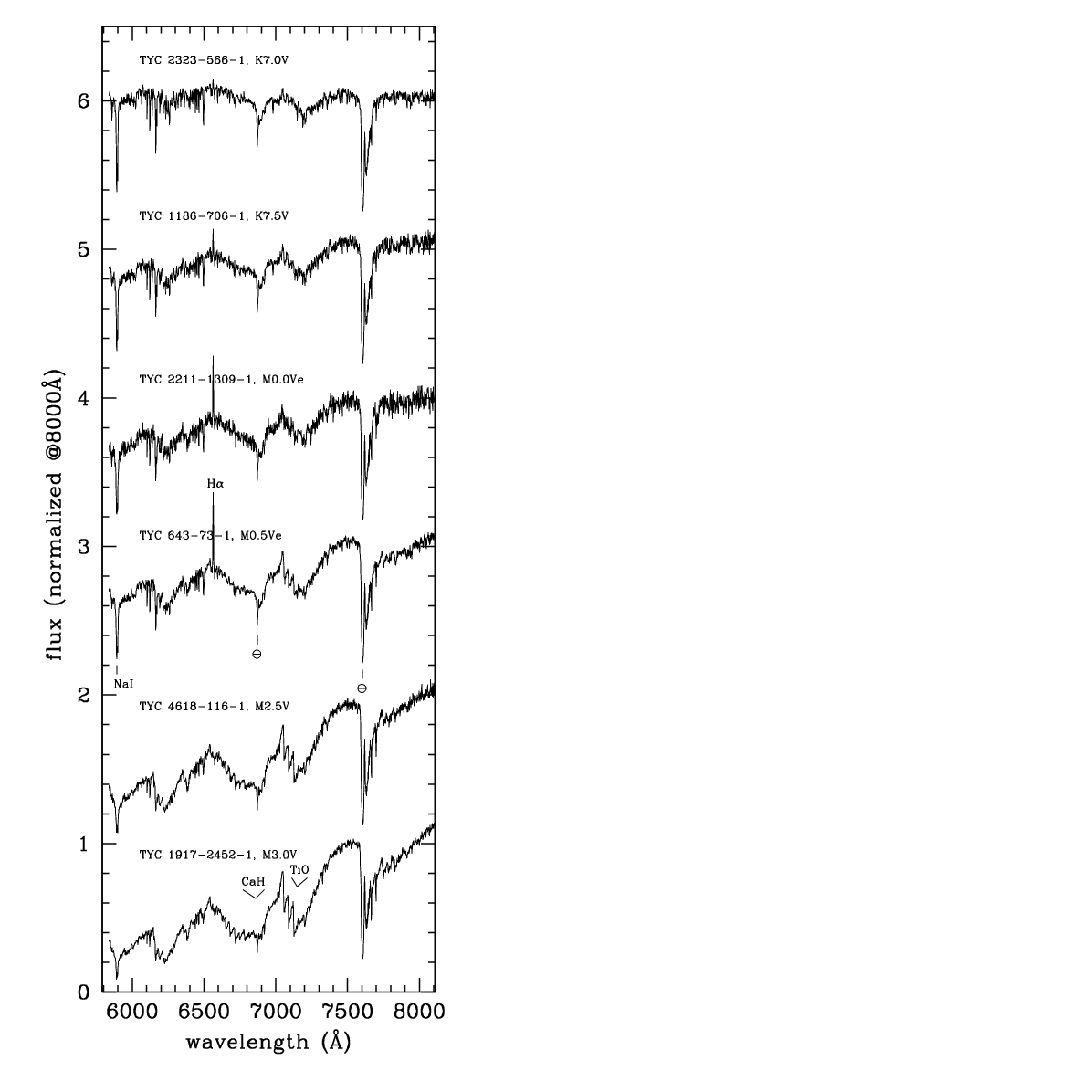}
\caption{Sample medium-resolution spectra of candidate BPMG
  members. The classification, based on bandstrengths of molecular TiO
  and CaH, is accurate to 0.5 subtype for late-K and M dwarfs. Some
  stars clearly show  H$\alpha$ in emission, which indicates
  chromospheric activity. The line is particularly strong in two of
  the most probable new members of the BPMG: TYC 2211-1309-1 (M0.0e)
  and TYC 643-73-1 (M0.5e). All spectra are normalized at 8000\AA\ and
  shifted vertically by integer values for clarity.}
\end{figure}

In the southern sky, spectra were obtained using the 1.5 m
telescope operated by the SMARTS consortium at the Cerro Tololo
Inter-American Observatory (CTIO). The telescope was equipped with the
R-C Cassegrain Spectrograph. We used Grating \#47 (800 lines mm$^{-1}$
blazed at 8000\AA) operated in the first order which, with the
110$\mu$m slit, provides a resolution of 3.1\AA. All the spectra were
obtained by service observers (along with calibration arcs and
spectrophotometric standards) and kindly processed by F. Walter using
the pipeline procedures he wrote and described in \citet{Wetal04}.

Spectra were obtained for 31 of the stars under investigation; a
subsample is displayed in Figure 5. Spectral types were also recovered
from the literature for TYC 3762-1492-1 (=EG Cam) and TYC 1942-2581-1 (=HO
Cnc), though both stars were rejected a priori as possible
members because of their {\it Hipparcos} parallax clearly at odds with the
predicted moving group distance. Spectral types were determined for
all the objects after comparison with spectra from classification
standards. Three stars were classified as G dwarfs, all others were
found to be either late-type K or early-type M dwarfs. For the M
dwarfs, subtypes were calculated based on the strength of the CaH and
TiO molecular bands around $7000\AA$, as measured by the TiO5, CaH2, and CaH3 molecular
band indices, and following the calibration of \citet{LRS03}. Subtypes
for the M dwarfs are all accurate to $\pm0.5$. Subtype
assignment for the G and K dwarfs is based on a comparison with
spectra from classification standards. Classification of G and K stars
using red spectra is however difficult, and some subtypes are accurate
to 1-2 subclasses; uncertain subtypes are noted with a colon (``:'').
Spectral classification can be found in the literature for some of
these stars; the subtypes generally agree with ours to $\pm1$
subtype. Many of the former spectral types are based of photographic
images from objective prism surveys \citep{Stephenson86,Stephenson86b}
some of which are off by more than 1 subtype; in these cases we deem
our CCD spectra to be more reliable. Assigned spectral types appear in
Table 2.

H$\alpha$ in emission was positively identified in seven
stars. The equivalent widths of the H$\alpha$ lines were calculated in
IRAF, using the {\tt onedspec} package. The results are shown in Table
2; stars for which we found no evidence of H$\alpha$ in emission are
noted as ``n.d.'' (for ``nondetection''). We rejected as possible
BPMG member any star showing no significant H$\alpha$ in emission {\em
  and} having no counterpart in {\it ROSAT}. This reduced the list
of possible candidates to only six stars, which were targeted for
follow-up radial velocity measurement.

\subsection{Radial velocity confirmation}

Radial velocity observations were obtained using the Cryogenic Echelle
Spectrograph (CSHELL), the high-resolution infrared
spectrometer at the NASA 3 m Infrared Telescope Facility (IRTF). The
detector is a 256$\times$256 InSb array. We set the
grating angle to provide 1.5548$\mu$m at the center of the array. At
this wavelength, CSHELL's free spectral range is $\sim$730 km
s$^{-1}$ (38 \AA). We used the 0.$\arcsec$5 slit for a resolution
R$\simeq$30,000. The instrumental setup was set by the requirements of
the part of our observing program directed toward pre-main-sequence
spectroscopic binaries \citep{SP04} but they are equally well suited
for the radial velocity measurements reported here. The spectra were
taken as a series of beam-switched, 300 s integrations and were
extracted using procedures described in \citet{BS08}. The radial
velocities were measured by cross-correlation using the same templates
as described in \citet{Metal02} and \citet{P02}.

We observed six known members of the BPMG in Table 1, and six
candidate members from the list compiled in Table 2. Radial velocities
were measured and compared to the radial velocities $V_{rmg}$
predicted assuming BPMG membership; results are compiled in Table
3. 

All six previously reported BPMG members have predicted radial
velocities $V_{rmg}$ within 4.5 km s$^{-1}$ of our observed
values, corroborating their BPMG membership. Three of the stars have
radial velocity measurements quoted in the literature. TYC 85-1075-1
(=V1005 Ori) is cited by \citet{joy48} to have a radial velocity of
+39 km s$^{-1}$; however this measurement was based on low-dispersion
spectrograms and is much less reliable than our own estimate of
+16.6$\pm$1.0 km s$^{-1}$. The star TYC 7460-137-1 (= AT Mic), was
measured by \citet{joy47} to have a radial velocity V$_{r}$=-4$\pm$2.4 km
s$^{-1}$, this time using higher dispersion spectrograms. The reported
value is largely consistent with our own (-5.5$\pm$1.5 km s$^{-1}$). The faint
companion to AT Mic, the star NLTT 49691, is quoted in the literature with
having a radial velocity V$_{r}$=-3$\pm$2.8 km s$^{-1}$
\citep{Montes01} which is also consistent with our own measurement
(-3.2$\pm$1.5 km s$^{-1}$). The close pair is part of a hierarchical
triple with TYC 7457-641-1 (=AU Mic). Several radial velocity
measurements exist for AU Mic; the most recent documented one is from
\citet{wilson67}, quoting V$_{r}$=+1.6$\pm$3.7 km s$^{-1}$ which is
only slightly off our own measured value.

For candidate members, we find that three stars (TYC 777-141-1, TYC
6806-631-1, and TYC 6849-1795-1) have predicted radial velocities
which are inconsistent with the measurements, with differences larger
than 15 km s$^{-1}$; this indicates that the three stars are
interlopers, whose proper motions are only aligned with the projected
motion of the BPMG by chance. The other three candidates (TYC
1186-706-1, TYC 7443-1102-1, and TYC 2211-1309-1) have predicted
radial velocities in excellent agreement with the observed values, all
within 3.5 km s$^{-1}$, comparable with the small differences observed
for the previously known members. All three stars are hereby
identified as highly probable members of the BPMG. All three stars
also show H$\alpha$ in emission and are X-ray emitters, which suggest they
are relatively young and active. We predict that parallax measurements
should confirm the distance estimates listed above, as an ultimate
test of moving group membership.

For all the stars observed at IRTF, we also estimate the projected
rotational velocities $v \sin{i}$ based of the width of the
atomic lines: cross-correlation is optimized using a suite of
slow-rotator templates, which are ``spun up'' using an algorithm
based on \cite{gray1992}. The templates are sampled every
2 km s$^{-1}$ (which defines the uncertainty on $v \sin{i}$). Results
are noted in Table 3. Of the three new members, only
one (TYC 2211-1309-1) is a moderately fast rotator ($v
sin{i}$=30$\pm$2 km s$^{-1}$). The other two have $v \sin{i} \lesssim$
7$\pm$2 km s$^{-1}$, which suggest either that they are not fast
rotators or that the rotation axis makes a small angle with the
line of sight. A slow rotation in itself would not, however, rule out
the star as a BPMG member. As one can see from Table 3, our
measurements indicate that several of the known BPMG members also have
relatively small values of $v \sin{i}$, with TYC 5832-666-1
(=BD-13$^{\circ}$6424) having $v \sin{i}$=7$\pm$2 km s$^{-1}$.

\subsection{Notes on the new BPMG members}

\subsubsection{TYC 1186-706-1}

The star TYC 1186-706-1, which we identify as a K7.5V dwarf, has a
predicted distance of 59.7pc. The star was first identified as a red
dwarf in the objective prism survey of \citet{Stephenson86}, based on
the strength of its sodium D line; it is star StKM 1\--34 in the
Stephenson catalog. The star was classified as K5, from a visual
inspection of the photographic plate spectrogram. This subtype is
broadly consistent with our own classification, although we believe
our CCD classification should be considered more reliable.

The star was also identified by \citet{Zickgraf03} to be the optical
counterpart to the X-ray source 1RXS J002334.9+201418 from the {\it ROSAT}
Bright Source Catalogue (RASS-BSC) of \citet{Voges99}, but the star
could not be classified because it was too bright and was saturated
on the Schmidt plates used in this objective prism survey.

More recently, the star was observed as part of the SuperWASP wide
field photometric survey, and discovered to be periodically variable
\citep{Norton07}. The star is cataloged under the name 1SWASP
J002334.66+201428.6, and is found to be variable at the 2\%\--3\% level
with a period of 7.9 days. Such variation is consistent with
rotational modulation.

\subsubsection{TYC 7443-1102-1}

The star TYC 7443-1102-1 is an M0.0V dwarf, at a predicted distance of
57.7pc. The predicted distance is similar to that of TYC 1186-706-1,
but the two stars are in very different directions on the sky (see
Figure 1) and must be physically unrelated. TYC 7443-1102-1 does not have
any previous mention in the literature. However, the common proper
motion companion identified while searching for X-ray counterparts
(see Section 4.2 above) is, of course, the known X-ray source
J195602.8-320720 \citep{Voges99}. We associate this X-ray source with
the proper motion star USNO-B1.0 0578-1079977. Its position is listed
in Table 3. 

The companion has a visual magnitude V$\approx$13.6 mag (estimated
from photographic plate measurements), compared with V$_T$=11.95 mag for
TYC 7443-1102-1. The companion should also be considered a new member
of the $\beta$ Pic moving group.

\subsubsection{TYC 2211-1309-1}

The fourth new member, TYC 2211-1309-1, is an M0.0Ve red dwarf at a
predicted distance of 45.6 pc. This object was first prosaically
identified as a ``noncluster X-ray source (Star?)'' by
\citet{Bohringer00}, in a search for galaxy clusters among {\it ROSAT}
sources. The object was identified by \citet{fuhrmeister03} as a
variable X-ray source with flaring behavior. This would be consistent
with chromospheric activity on that star.

This star was also identified in the SuperWASP wide field photometric
survey of \citet{Norton07} and found to be variable in the
optical regime. The star is cataloged under the name 1SWASP
J220041.59+271513.5, and was observed to be variable at the 2\% level,
with a period of 0.52 days. This behavior suggests rotational
modulation in a fast rotator, and is consistent with the high value of
$v \sin{i}$ we infer for the rotational velocity. An early M-type star
with a rotation period of 0.5 days, having
a radius of about 0.5 R$_{\odot}$ \citep{berger06}, would be expected
to have a rotation velocity at the equator of 50 km s$^{-1}$,
consistent with our inferred value of $v \sin{i}$ = 30 km s$^{-1}$.

\section{Conclusions}

We have presented an algorithm to identify probable members of any
known nearby moving group from large catalogs of proper
motion stars. The technique identifies prospective members based on the
orientation of their proper motion vector, which aligns with the
projected motion of the group in the plane of the sky, and uses the
magnitude of the proper motion to estimate a distance of the
assumption of group membership. That predicted distance is used to
verify consistency with the known color\--magnitude relationship of the
group.

We have tested the method by conducting a search for possible members
of the $\beta$ Pictoris moving group in a subsample of high proper
motion stars $\mu>70$ mas$^{-1}$ from the Tycho-2 catalog. A subset of
31 candidates was investigated for signs of youth, including
low-resolution spectroscopy to look for H$\alpha$ lines in emission,
and a search for X-ray counterparts in the {\it ROSAT} catalogs. Twelve
stars showing possible evidence for youth were targeted for
high-resolution infrared spectroscopy, in order to corroborate moving
group membership from their radial velocities. Three of the stars are
found to be convincing new members of the BPMG, these are TYC
1186-706-1, TYC 7443-1102-1, and TYC 2211-1309-1. The star TYC
7443-1102-1 is also found to have a common proper motion companion,
with an angular separation of 26$\arcsec$, which is most probably a
wide companion with a projected separation $\approx$1,200 AU.

We believe that the method can be successfully expanded to the
identification of members from any other moving group and
association. The use of a deeper proper motion catalog, such as
the recent PPM-Extended (PPMX) catalog \citep{Roser08}, which extends
to V=15 mag, or the SUPERBLINK database \citep{LS05} that is
statistically complete to V=19 mag, should allow for the
identification of most low-mass members of these moving groups, at
least down to the hydrogen burning limit. 

\acknowledgments

This research has been supported by NSF grant AST-0607757 at
the American Museum of Natural History. At Stony Brook, the research
was supported by JPL contract 125094, and by NSF Grant 06-07612.

\clearpage
\begin{landscape}
\begin{deluxetable}{rrlrrrrrrrrrrr}
\tabletypesize{\scriptsize}
\tablecaption{Recovered $\beta$ Pictoris moving group members}
\tablehead{
\colhead{TYC} & 
\colhead{HIP} & 
\colhead{Alt. name} & 
\colhead{$\alpha$(ICRS)\tablenotemark{a}} & 
\colhead{$\delta$(ICRS)\tablenotemark{a}} & 
\colhead{$\mu_{\alpha}$} & 
\colhead{$\mu_{\delta}$} & 
\colhead{V$_T$} & 
\colhead{K$_{\rm s}$} & 
\colhead{d$_{\rm kin}$\tablenotemark{b}} & 
\colhead{d$_{\rm hip}$\tablenotemark{c}} & 
\colhead{X-ray CR\tablenotemark{e}} & 
\colhead{$HR1$\tablenotemark{f}} & 
\colhead{$HR2$\tablenotemark{f}} \\
\colhead{} & 
\colhead{} & 
\colhead{} & 
\colhead{(2000.0)\tablenotemark{d}} & 
\colhead{(2000.0)\tablenotemark{d}} & 
\colhead{mas yr$^{-1}$} & 
\colhead{mas yr$^{-1}$} & 
\colhead{mag} & 
\colhead{mag} & 
\colhead{pc} & 
\colhead{pc} & 
\colhead{ks$^{-1}$} &  
\colhead{} & 
\colhead{}
}
\startdata
6412 1068 1&     560&      HD 203&   1.708454& -23.107429&   97.1&  -47.4&  6.18& 5.24& 39.4$\pm$0.3&   39.4$\pm$0.6&   44$\pm$15&  0.32$\pm$0.32& -0.30$\pm$0.37\\
1777 1480 1&   10679&BD+28$^{\circ}$382B&  34.352851&  28.741941&   86.7&  -76.7&  7.76& 6.26& 37.4$\pm$0.9&   27.3$\pm$4.4&  364$\pm$73& -0.03$\pm$0.20&  0.13$\pm$0.28\\
1777 1479 1&   10680&    HD 14082&  34.355136&  28.745208&   84.3&  -77.6&  7.04& 5.78& 37.8$\pm$0.7&   34.5$\pm$3.4&     \nodata&\nodata&\nodata\\
  45  990 1&   11360&    HD 15115&  36.567489&   6.292664&  100.7&    5.5&  6.79& 5.82& 38.7$\pm$0.5&   45.2$\pm$1.3&   77$\pm$20&  0.04$\pm$0.25& -0.61$\pm$0.55\\
2323  566 1&   11437&      AG Tri&  36.871692&  30.973667&   84.0&  -71.8& 10.08& 7.08& 39.2$\pm$1.1&   40.0$\pm$3.6&  566$\pm$46& -0.16$\pm$0.07& -0.17$\pm$0.12\\
4739 1551 1&   21547&       c Eri&  69.400451&  -2.473389&   43.6&  -64.1&  5.21& 4.53& 32.2$\pm$0.4&   29.4$\pm$0.3&     \nodata&\nodata&\nodata\\
  85 1075 1&   23200&   V1005 Ori&  74.895042&   1.783751&   38.1&  -94.4& 10.10& 6.26& 25.5$\pm$0.5&   25.9$\pm$1.7&  651$\pm$42& -0.19$\pm$0.06&  0.28$\pm$0.10\\
8513  572 1&   23309&CD-57$^{\circ}$1054&  75.196235& -57.257244&   36.2&   72.6& 10.11& 6.24& 24.3$\pm$0.7&   26.8$\pm$0.8&  330$\pm$52&  0.11$\pm$0.15& -0.15$\pm$0.18\\
 689  130 3&   23418&  [RHG95]853&  75.494987&   9.983299&   17.2&  -82.0& 11.74& 6.37& 36.4$\pm$4.8&  33.2$\pm$10.5& 661$\pm$59& -0.34$\pm$0.08& -0.07$\pm$0.14\\
8099 1392 1&   27321& $\beta$ Pic&  86.821182& -51.066703&    4.1&   83.3&  3.85& 3.52& 17.1$\pm$0.3&   19.4$\pm$0.1&     \nodata&\nodata&\nodata\\
9172  690 1&   29964&      AO Men&  94.617607& -72.045021&   -8.5&   75.7&  9.99& 6.81& 37.2$\pm$0.7&   38.6$\pm$1.3& 1030$\pm$30& -0.07$\pm$0.02&  0.03$\pm$0.04\\
8704 1271 1&   76629&    V343 Nor& 234.740005& -57.707344&  -46.2&  -97.9&  8.16& 5.85& 40.6$\pm$0.6&   39.8$\pm$1.7& 1420$\pm$72& -0.07$\pm$0.05&  0.13$\pm$0.07\\
6805 1909 1&   79881&       d Sco& 244.574661& -28.613777&  -32.4& -100.2&  4.78& 4.73& 37.9$\pm$0.3&   41.3$\pm$0.4&     \nodata&\nodata&\nodata\\
9064 3514 1&   84586&    V824 Ara& 259.356384& -66.950714&  -21.6& -136.4&  6.87& 4.70& 31.5$\pm$0.3&   31.4$\pm$0.5&  476$\pm$27& -0.06$\pm$0.05&  0.02$\pm$0.08\\
8369 1619 1&   88399&   HD 164249& 270.764190& -51.648807&    2.8&  -87.2&  7.01& 5.91& 50.2$\pm$0.6&   48.1$\pm$1.3&  150$\pm$42&  0.13$\pm$0.29&  0.36$\pm$0.36\\
7911 5035 1&   88726&   HD 165189& 271.707855& -43.424964&   13.8& -105.3&  5.63& 4.38& 40.2$\pm$0.9&   41.8$\pm$1.2&     \nodata&\nodata&\nodata\\
9077 2487 1&   92024&   HD 172555& 281.361907& -64.870918&   32.9& -148.2&  4.77& 4.29& 28.8$\pm$0.2&   28.5$\pm$0.2&   97$\pm$31& -0.22$\pm$0.33& -0.32$\pm$0.67\\
9073  762 1& \nodata&     \nodata& 281.718963& -62.176612&   18.1&  -76.6& 12.22& 7.85& 55.8$\pm$3.5&\nodata        &  137$\pm$38& -0.22$\pm$0.29&  0.38$\pm$0.43\\
7408   54 1& \nodata&     \nodata& 282.685302& -31.796323&   10.6&  -77.8& 11.30& 7.46& 51.1$\pm$1.9&\nodata        &  308$\pm$37& -0.19$\pm$0.11&  0.07$\pm$0.19\\
8381 2435 1&   92680&      PZ Tel& 283.274414& -50.180332&   15.8&  -84.1&  8.41& 6.36& 51.2$\pm$0.7&   51.5$\pm$2.6& 1000$\pm$88&  0.02$\pm$0.08&  0.10$\pm$0.12\\
8765 2571 1&   95261&   HD 181296& 290.713256& -54.423740&   25.0&  -83.1&  5.02& 5.00& 50.8$\pm$0.7&   48.2$\pm$0.5&     \nodata&\nodata&\nodata\\
8765  638 1&   95270&   HD 181327& 290.745513& -54.537860&   24.1&  -82.9&  7.04& 5.91& 51.1$\pm$0.9&   51.8$\pm$1.7&     \nodata&\nodata&\nodata\\
6909 1892 1&   99273&   HD 191089& 302.271636& -26.223878&  -52.4&  -58.8&  7.19& 6.07& 51.6$\pm$0.9&   52.2$\pm$1.2&   73$\pm$18& -0.30$\pm$0.22&  0.07$\pm$0.40\\
7460  137 1&  102141&      AT Mic& 310.462432& -32.434368&  261.3& -344.8& 11.24& 4.94&  9.8$\pm$0.1&   10.7$\pm$0.4&3910$\pm$120& -0.19$\pm$0.03&  0.06$\pm$0.04\\
7457  641 1&  102409&      AU Mic& 311.288940& -31.340036&  278.8& -360.0&  8.77& 4.52&  9.3$\pm$0.1&    9.9$\pm$0.1&5950$\pm$121& -0.07$\pm$0.02&  0.06$\pm$0.03\\
6348   98 1&  103311&   HD 199143& 313.948486& -17.114032&   62.2&  -65.4&  7.32& 5.81& 44.6$\pm$1.0&   45.7$\pm$1.6&1400$\pm$107&  0.24$\pm$0.07&  0.10$\pm$0.09\\
6349  200 1& \nodata&      AZ Cap& 314.011260& -17.181446&   59.3&  -63.0& 10.63& 7.03& 46.6$\pm$2.3&\nodata        &  236$\pm$26& -0.23$\pm$0.10&  0.01$\pm$0.17\\
9340  437 1& \nodata&CPD-72$^{\circ}$2713& 340.703155& -71.705772&   94.1&  -54.4& 10.54& 6.89& 36.7$\pm$0.7&\nodata        &  727$\pm$78& -0.21$\pm$0.10&  0.27$\pm$0.16\\
5832  666 1& \nodata& BD-13$^{\circ}$6424& 353.128265& -12.264112&  138.1&  -83.2& 10.71& 6.56& 27.3$\pm$0.4&\nodata        &  620$\pm$48& -0.10$\pm$0.07&  0.10$\pm$0.11\\
\cutinhead{TW Hya association (TWA) interlopers}		 		   			      
7190 2111 1& \nodata&     \nodata& 160.625610& -33.671215& -122.2&  -29.3& 10.92& 6.89& 30.2$\pm$0.6&\nodata        &  324$\pm$31& -0.08$\pm$0.09&  0.05$\pm$0.14\\
7208  347 1&   53911&      TW Hya& 165.466476& -34.704719&  -73.4&  -17.5& 11.27& 7.29& 52.0$\pm$1.6&   53.7$\pm$6.2&  571$\pm$43&  0.58$\pm$0.06& -0.12$\pm$0.08\\
7201   27 1& \nodata&     \nodata& 167.307785& -30.027683&  -90.0&  -87.7& 11.16& 6.71& 43.6$\pm$2.0&\nodata        &  341$\pm$35& -0.22$\pm$0.09& -0.02$\pm$0.15\\
7760  283 1& \nodata&     \nodata& 183.878219& -39.811759&  -75.9&  -26.3& 11.43& 7.30& 52.8$\pm$1.6&\nodata        &  375$\pm$65& -0.31$\pm$0.16&  0.31$\pm$0.27\\
\enddata
\tablenotetext{a}{Right ascension ($\alpha$) and declinations
  ($\delta$) are given in the International Celestial Reference System
  (ICRS) and at the epoch 2000.0, as provided by the Tycho-2 catalog.}
\tablenotetext{b}{Kinematic distance based on proper motion and
  assumed group membership.}
\tablenotetext{c}{Geometric distance based on the Hipparcos
  parallax.}
\tablenotetext{d}{Epoch of the ICRS coordinates.}
\tablenotetext{e}{{\it ROSAT} X-ray count rate.}
\tablenotetext{f}{{\it ROSAT} X-ray hardness ratio.}
\end{deluxetable}
\clearpage
\end{landscape}

\clearpage
\begin{landscape}
\begin{deluxetable}{rrlrrrrrrrrrrrrrc}
\tabletypesize{\scriptsize}
\tablecaption{Candidate $\beta$ Pictoris moving group members}
\tablehead{
\colhead{TYC} & 
\colhead{HIP} & 
\colhead{Alt. name} & 
\colhead{$\alpha$(ICRS)\tablenotemark{a}} & 
\colhead{$\delta$(ICRS)\tablenotemark{a}} & 
\colhead{$\mu_{\alpha}$} & 
\colhead{$\mu_{\delta}$} & 
\colhead{V$_T$} & 
\colhead{K$_{\rm s}$} & 
\colhead{d$_{\rm kin}$\tablenotemark{b}} & 
\colhead{d$_{\rm hip}$\tablenotemark{c}} & 
\colhead{X-ray CR\tablenotemark{e}} & 
\colhead{$HR1$\tablenotemark{f}} & 
\colhead{$HR2$\tablenotemark{f}} &
\colhead{Sp.} &
\colhead{EW(H$\alpha$)} &
\colhead{BPMG} \\
\colhead{} & 
\colhead{} & 
\colhead{} & 
\colhead{(2000.0)\tablenotemark{d}} & 
\colhead{(2000.0)\tablenotemark{d}} & 
\colhead{mas yr$^{-1}$} & 
\colhead{mas yr$^{-1}$} & 
\colhead{mag} & 
\colhead{mag} & 
\colhead{pc} & 
\colhead{pc} & 
\colhead{ks$^{-1}$} & 
\colhead{} & 
\colhead{} & 
\colhead{type} & 
\colhead{$\AA$} &
\colhead{member?}
}
\startdata
1186  706 1& \nodata&    \nodata&           5.894279&  20.241407&   63.0&  -38.1& 10.85& 7.33&  59.7$\pm$1.6&      \nodata& 176$\pm$27& -0.12$\pm$0.15& -0.07$\pm$0.24& K7.5V& -0.7$\pm$0.1  & table 3\\ 
5853  933 1& \nodata&  BPM 47010&          18.914964& -21.514326&   81.4&  -22.5& 13.03& 8.52&  45.9$\pm$2.0&      \nodata&    \nodata&        \nodata&        \nodata& K7.0V&         n.d.  & no\\
 643   73 1&   12787&    \nodata&          41.088851&  10.961535&   73.5&  -57.3& 11.40& 7.11&  41.9$\pm$1.2& 34.9$\pm$3.7& 301$\pm$94& -0.33$\pm$0.32& -0.40$\pm$0.94& M0.5Ve& -3.4$\pm$0.1 & no\\ 
3333 1029 1& \nodata&    \nodata&          64.939537&  47.758914&   21.2&  -79.0& 11.32& 7.69&  53.5$\pm$3.1&      \nodata&    \nodata&        \nodata&        \nodata& M0.0V&         n.d.  & no\\
2384 1106 1& \nodata&  HD 279890&          66.975112&  35.936912&   29.8&  -64.5& 10.82& 7.20&  58.7$\pm$4.1&      \nodata&    \nodata&        \nodata&        \nodata&  K0V:&         n.d.  & no\\
3762 1492 1&   28368&     EG Cam&          89.907287&  58.593612&   11.2& -252.9& 10.27& 6.21&  17.4$\pm$0.2& 13.5$\pm$0.3&    \nodata&        \nodata&        \nodata& M0.5V\tablenotemark{g}&    \nodata  & no\\ 
8895  225 1&   31878&CD-61$^{\circ}$1439&  99.958557& -61.478378&  -25.7&   71.3&  9.69& 6.50&  29.3$\pm$1.0& 22.4$\pm$0.5& 143$\pm$14& -0.05$\pm$0.09& -0.13$\pm$0.13&  K5V:&         n.d.  & no\\ 
2443  845 1& \nodata& StKM 1-603&         100.331420&  33.907360&  -25.6&  -68.1& 11.33& 7.96&  55.8$\pm$2.8&      \nodata&    \nodata&        \nodata&        \nodata& M0.0V&         n.d.  & no\\
1903 1306 1&   34222&    GJ 265A&         106.426132&  27.471057&  -49.2&  -97.2& 10.26& 6.78&  35.8$\pm$0.7& 23.9$\pm$1.3&    \nodata&        \nodata&        \nodata& K7.0V&         n.d.  & no\\ 
1917 2452 1&   35191&    GJ 9227&         109.082481&  27.142967&  -39.0& -191.6& 10.75& 6.18&  20.0$\pm$0.6& 12.0$\pm$0.3&    \nodata&        \nodata&        \nodata& M3.0V&         n.d.  & no\\ 
4618  116 1& \nodata&    \nodata&         109.992683&  84.077461&  -38.5&  -86.0& 11.93& 7.51&  41.8$\pm$1.3&      \nodata&    \nodata&        \nodata&        \nodata& M2.5V&         n.d.  & no\\
 777  141 1& \nodata&    \nodata&         113.734748&  14.765309&  -80.1& -106.4& 10.74& 6.39&  26.4$\pm$0.7&      \nodata&    \nodata&        \nodata&        \nodata& M3.0Ve& -3.3$\pm$0.1 & table 3\\ 
9389   53 1& \nodata&    \nodata&         117.695777& -79.867774&  -48.0&   57.7& 11.18& 7.81&  44.5$\pm$2.1&      \nodata&    \nodata&        \nodata&        \nodata&  K3V:&       n.d.  & no\\
1942 2581 1&   42253&     HO Cnc&         129.232696&  23.246904& -108.7& -106.7&  9.51& 6.41&  26.8$\pm$0.4& 44.1$\pm$2.7&    \nodata&        \nodata&        \nodata&   K5V\tablenotemark{h}&      \nodata  & no\\ 
1952 1263 1& \nodata& StKM 1-764&         139.756805&  23.597393&  -49.6&  -55.0& 11.92& 8.36&  57.4$\pm$1.8&      \nodata&    \nodata&        \nodata&        \nodata& K7.5V&         n.d.  & no\\
 240 2164 1&   48447&    \nodata&         148.163299&   3.130231&  -61.6&  -33.7& 10.56& 7.07&  57.8$\pm$1.5& 22.7$\pm$1.4&    \nodata&        \nodata&        \nodata& K7.5V&         n.d.  & no\\ 
8606  821 1& \nodata&    \nodata&         149.721710& -57.824100& -165.9&   36.5& 11.88& 6.44&  19.4$\pm$0.5&      \nodata&    \nodata&        \nodata&        \nodata& M4.5V&         n.d.  & no\\
7732 1096 1& \nodata&    \nodata&         164.767608& -41.300849& -118.8&  -75.4& 11.84& 7.12&  27.2$\pm$0.7&      \nodata&    \nodata&        \nodata&        \nodata& M1.5V&         n.d.  & no\\
7322   67 1& \nodata&   LTT 6184&         232.204833& -34.332885& -104.2& -118.5&  9.04& 5.77&  26.9$\pm$0.3&      \nodata&    \nodata&        \nodata&        \nodata&  G9V:&         n.d.  & no\\
8304 1011 1& \nodata&    \nodata&         233.453536& -49.300254& -138.3& -213.6& 11.16& 6.47&  17.3$\pm$0.3&      \nodata&    \nodata&        \nodata&        \nodata&  K3V:&         n.d.  & no\\
6778  560 1& \nodata&BD-22$^{\circ}$4030& 237.258209& -23.042593&  -51.5&  -74.5& 10.01& 6.89&  43.2$\pm$1.2&      \nodata&    \nodata&        \nodata&        \nodata&  G5V:&         n.d.  & no\\
5046  481 1& \nodata&BD-05$^{\circ}$4273& 244.614517&  -6.074768&  -32.5&  -78.8& 10.05& 6.50&  36.1$\pm$1.1&      \nodata&    \nodata&        \nodata&        \nodata&  K0V:&        n.d.  & no\\
1513  989 1& \nodata&  HD 147926&         245.998031&  20.394264&  -63.6&  -47.1&  8.13& 4.96&  21.7$\pm$0.3&      \nodata&    \nodata&        \nodata&        \nodata&  K2V:&        n.d.  & no\\
6806  631 1& \nodata&  HD 147611&         246.008605& -29.178794&  -42.3& -139.5& 10.15& 6.77&  27.4$\pm$0.5&      \nodata& 505$\pm$39& -0.02$\pm$0.07& -0.07$\pm$0.11&  K5V:& -0.3$\pm$0.1 & table 3\\ 
7892 2635 1& \nodata&    \nodata&         264.482879& -42.324363&  -11.6& -107.2& 11.60& 7.45&  39.4$\pm$1.2&      \nodata&    \nodata&        \nodata&        \nodata& M1.5V&        n.d.  & no\\
6832  549 1& \nodata&    \nodata&         266.221771& -25.644212&  -14.7&  -82.5& 11.03& 7.49&  44.5$\pm$2.4&      \nodata&    \nodata&        \nodata&        \nodata&  K2V:&        n.d.  & no\\
6849 1795 1& \nodata&    \nodata&         268.725524& -26.827995&   18.1& -103.5& 10.28& 6.91&  36.0$\pm$1.0&      \nodata& 563$\pm$45& -0.10$\pm$0.08&  0.02$\pm$0.12&  K5V:& -1.1$\pm$0.1 & table 3\\ 
6262 1013 1& \nodata&    \nodata&         268.909790& -22.320178&   -9.6&  -70.4& 11.71& 7.75&  50.5$\pm$3.2&      \nodata&    \nodata&        \nodata&        \nodata&  K0V:&        n.d.  & no\\
7899 6490 1& \nodata&  HD 324810&         270.559997& -37.682273&   18.9& -185.5& 11.95& 6.97&  22.2$\pm$0.5&      \nodata&    \nodata&        \nodata&        \nodata&\nodata&        n.d.  & no\\
7443 1102 1& \nodata&    \nodata&         299.018127& -32.126995&   31.2&  -65.0& 11.80& 7.84&  57.7$\pm$2.8&      \nodata&    \nodata&        \nodata&        \nodata& M0.0V& -0.4$\pm$0.1 & table 3\\ 
2211 1309 1& \nodata&    \nodata&         330.173126&  27.253810&   76.3&  -14.7& 11.37& 7.72&  45.6$\pm$1.6&      \nodata& 219$\pm$22& -0.20$\pm$0.09& -0.23$\pm$0.15& M0.0Ve& -2.6$\pm$0.4 & table 3\\ 
4275 1683 1& \nodata&    \nodata&         330.877685&  66.683044&   72.6&   14.5& 11.19& 7.85&  45.6$\pm$1.6&      \nodata&    \nodata&        \nodata&        \nodata& K7.5V&         n.d.  & no\\
4285 2546 1& \nodata&    \nodata&         357.159545&  62.026763&   77.9&   -8.6& 11.34& 7.58&  49.5$\pm$2.2&      \nodata&    \nodata&        \nodata&        \nodata& K5.0V&         n.d.  & no\\
\enddata
\tablenotetext{a}{Right ascension ($\alpha$) and declinations
  ($\delta$) are given in the International Celestial Reference System
  (ICRS) and at the epoch 2000.0, as provided by the Tycho-2 catalog.}
\tablenotetext{b}{Kinematic distance based on proper motion and
  assumed group membership.}
\tablenotetext{c}{Geometric distance based on the Hipparcos
  parallax.}
\tablenotetext{d}{Epoch of the ICRS coordinates.}
\tablenotetext{e}{{\it ROSAT} X-ray count rate.}
\tablenotetext{f}{{\it ROSAT} X-ray hardness ratio.}
\tablenotetext{g}{Spectral type from \citet{reid95}}
\tablenotetext{h}{Spectral type from \citet{Montes01}}
\end{deluxetable}
\clearpage
\end{landscape}

\begin{deluxetable*}{rlrrrrrcr}
\tabletypesize{\scriptsize}
\tablewidth{450pt}
\tablecaption{Results from high-resolution infrared spectroscopy}
\tablehead{
\colhead{TYC} & 
\colhead{Alt. name} & 
\colhead{$\alpha$(ICRS)} & 
\colhead{$\delta$(ICRS)} &
\colhead{V$_{rmg}$\tablenotemark{a}} & 
\colhead{V$_{r}$\tablenotemark{b}} &
\colhead{$v \sin{i}$\tablenotemark{c}} &
\colhead{Sp.}&
\colhead{BPMG}\\
\colhead{} & 
\colhead{} & 
\colhead{(2000.0)} & 
\colhead{(2000.0)} & 
\colhead{(km s$^{-1}$)} & 
\colhead{(km s$^{-1}$)} &
\colhead{(km s$^{-1}$)} &
\colhead{type}&
\colhead{member?}
}
\startdata
\cutinhead{Previously known BPMG members}
  85-1075-1& V1005 Ori&    74.895042&   1.783751& $+17.7$& $+16.6\pm1.0$&           12$\pm$2& M0.5Ve & yes\\
  7408-54-1&   \nodata&   282.685302& -31.796323&  $-9.6$&  $-7.1\pm1.5$&           50$\pm$2&   K7Ve & yes\\
7460-137-1&     AT Mic&   310.462432& -32.434368&  $-5.9$&  $-5.5\pm1.5$&           15$\pm$2& M4.0Ve & yes\\
\nodata& NLTT 49691\tablenotemark{d}& 310.463112& -32.436175& $-5.9\pm$& $-3.2\pm1.5$& 12$\pm$2& M4.0V  & yes\\
7457-641-1&     AU Mic&   311.288940& -31.340036&  $-6.0$&  $-6.7\pm1.5$&           12$\pm$2& M1.0Ve & yes\\
6349-200-1&     AZ Cap&   314.011260& -17.181446&  $-9.2$&  $-4.7\pm1.5$&           20$\pm$2&   K5Ve & yes\\
5832-666-1& BD-13$^{\circ}$6424& 353.128265& -12.264112& $+0.6$& $+0.5\pm1.5$&       7$\pm$2& M0.5Ve & yes\\
\cutinhead{New BPMG candidates}			       		   
 1186-706-1& \nodata&     5.894279&  20.241407&  $-1.5$&  $-5.0\pm1.5$&              7$\pm$2&  K7.5V & yes\\
  777-141-1& \nodata&   113.734748&  14.765309& $+13.5$& $+31.4\pm1.0$&              6$\pm$2&  M3.0Ve& no \\
 6806-631-1&HD 147611&  246.008605& -29.178794&  $-9.7$&  $+7.4\pm1.0$&              4$\pm$2&  K4V:  & no \\
6849-1795-1& \nodata&   268.725524& -26.827995& $-11.7$& $-28.4\pm1.5$&             25$\pm$2&  K4V:  & no \\
7443-1102-1& \nodata&   299.018127& -32.126995&  $-7.8$&  $-8.8\pm1.5$&              6$\pm$2&  M0.0V & yes\\
\nodata& 1RXS J195602.8-320720\tablenotemark{e}& 299.012262& -32.121850& \nodata& \nodata&  \nodata& \nodata& yes \\
2211-1309-1& \nodata&   330.173126&  27.253810& $-12.7$& $-13.3\pm2.4$&             30$\pm$2&  M0.0Ve& yes\\
\enddata
\tablenotetext{a}{Predicted radial velocity assuming the star to be a
  BPMG member.}
\tablenotetext{b}{Radial velocity measured from high-resolution
  infrared spectroscopy (IRTF).}
\tablenotetext{c}{Measured rotational velocity.}
\tablenotetext{d}{Resolved companion to TYC 7460-137-1, with angular
  separation $\rho=$3.349$\pm$0.007 \citep{}.}
\tablenotetext{d}{Common proper motion companion to TYC 7443-1102-1}
\end{deluxetable*}

\end{document}